\documentclass[sigconf,nonacm]{aamas} 

\usepackage{balance}
\usepackage{cleveref}
\usepackage{tikz}
\usepackage{pifont}
\usepackage{xspace}
\usepackage{listings}

\def\tool {IMITATOR4AMAS}

\usetikzlibrary{automata, arrows.meta, positioning,shapes.multipart,fit,calc,3d,perspective,arrows,decorations.pathreplacing,shapes.misc}
\definecolor{darkgreen}{RGB}{0,120,0}

\newcommand{\eg}{e.g.\xspace}
\newcommand{\ie}{i.e.\xspace}

\crefname{definition}{def.}{defs.}
\Crefname{definition}{Definition}{Definitions}
\crefname{example}{ex.}{exs.}
\Crefname{example}{Example}{Examples}
\crefname{algorithm}{alg.}{algs.}
\Crefname{algorithm}{Algorithm}{Algorithms}
\crefname{table}{tab.}{tabs.}
\Crefname{table}{Table}{Tables}
\crefname{figure}{fig.}{figs.}
\Crefname{figure}{Figure}{Figures}
\crefname{section}{Section}{Sections}
\Crefname{section}{Section}{Sections}
\crefname{subsection}{Section}{Sections}
\Crefname{subsection}{Section}{Sections}

\makeatletter
\renewcommand{\paragraph}{\@startsection{paragraph}{4}{0pt}%
  {.8ex plus 0.2ex minus 0.2ex}%
  {-0.5em}%
  {\bfseries}}
\makeatother

\makeatletter
\newcommand\dontbreakpar{\par\nobreak\@afterheading}
\makeatother

\def\event/{action}
\def\Event/{Action}

\newcommand{\TAMAS}{CAMAS\xspace}

\newcommand{\A}{\ensuremath{A}}

\newcommand{\prop}[1]{\ensuremath{\mathsf{{#1}}}}

\newcommand{\coop}[1]{\langle\!\langle{#1}\rangle\!\rangle}
\newcommand{\Epath}{\mathsf{\exists}}
\newcommand{\Apath}{\mathsf{\forall}}

\newcommand{\Sometm}{\mathtt{F}\,}
\newcommand{\Always}{\mathtt{G}\,}
\newcommand{\Until}{\,\mathtt{U}\,}
\newcommand{\Release}{\,\mathtt{R}\,}

\newcommand{\stratstyle}[1]{\ensuremath{\mathrm{#1}}}

\newcommand{\ir}{\stratstyle{ir}\xspace}

\newcommand{\lan}[1]{\ensuremath{\mathbf{#1}}\xspace}

\newcommand{\CTL}[1][]{\lan{CTL}}
\newcommand{\CTLK}[1][]{\lan{CTLK}}

\newcommand{\SLSG}{\lan{SL[SG]}}

\newcommand{\TCTL}[1][]{\lan{TCTL_{\stratstyle{#1}}}}

\newcommand{\MTL}[1][]{\lan{MTL}}
\newcommand{\SMTL}[1][]{\lan{SMTL}}
\newcommand{\MITL}[1][]{\lan{MITL}}
\newcommand{\TPTL}[1][]{\lan{TPTL}}

\newcommand{\STCTL}[1][]{\lan{STCTL_{\stratstyle{#1}}}}

\newcommand{\imitator}{IMITATOR\xspace}

\definecolor{coloract}{rgb}{0.50,0.70,0.30}
\definecolor{colorclock}{rgb}{0.4,0.4,1}
\definecolor{colorloc}{rgb}{0.4,0.4,0.65}
\definecolor{colorparam}{rgb}{1,0.6,0.0}
\definecolor{colordisc}{rgb}{1, 0, 1}
\definecolor{colorconst}{rgb}{0.50, 0.20, 0.00}

\tikzstyle{every node}=[initial text=]
\tikzstyle{location}=[rectangle, rounded corners, minimum size=12pt, draw=black, fill=blue!10, inner sep=2pt]
\tikzstyle{invariant}=[draw=black, dotted, inner sep=1pt] 
\tikzstyle{final}=[double]

\setcopyright{ifaamas}
\acmConference[AAMAS '26]{Proc.\@ of the 25th International Conference
on Autonomous Agents and Multiagent Systems (AAMAS 2026)}{May 25 -- 29, 2026}
{Paphos, Cyprus}{C.~Amato, L.~Dennis, V.~Mascardi, J.~Thangarajah (eds.)}
\copyrightyear{2026}
\acmYear{2026}
\acmDOI{}
\acmPrice{}
\acmISBN{}

\acmSubmissionID{0}

\title[\tool]{\tool: Strategy Synthesis for STCTL}

\author{Davide Catta}
\affiliation{
    \institution{LIPN, CNRS UMR 7030, \\ Université Sorbonne Paris Nord}
    \city{Villetaneuse}
    \country{France}
}
\email{davide.catta@lipn.univ-paris13.fr}

\author{Adrien Lacroix}
\affiliation{
  \institution{LIPN, CNRS UMR 7030, \\ Université Sorbonne Paris Nord}
	\country{}
  \institution{\& Ecole Centrale de Nantes, France}
   \country{}
  }
  \email{Adrien.Lacroix@eleves.ec-nantes.fr}

\author{Wojciech Penczek}
\affiliation{
  \institution{Institute of Computer Science, \\ Polish Academy of Sciences}
	\city{Warsaw}
  \country{Poland}
  }
  \email{penczek@ipipan.waw.pl}
  
\author{Laure Petrucci}
\affiliation{
    \institution{LIPN, CNRS UMR 7030, \\ Université Sorbonne Paris Nord}
    \city{Villetaneuse}
    \country{France}
}
\email{laure.petrucci@lipn.univ-paris13.fr}
  
\author{Teofil Sidoruk}
\affiliation{
  \institution{Institute of Computer Science, \\ Polish Academy of Sciences}
	\city{Warsaw}
  \country{Poland}
  }
  \email{t.sidoruk@ipipan.waw.pl} 

\begin{abstract}
\tool\ supports model checking and synthesis of memoryless imperfect information 
strategies for \STCTL, interpreted over networks of parametric timed automata 
with asynchronous execution.
While extending the verifier \imitator, \tool\ is the first tool 
for strategy synthesis in this setting.
Our experimental results show a substantial speedup over previous approaches.

\end{abstract}


\newcommand{\BibTeX}{\rm B\kern-.05em{\sc i\kern-.025em b}\kern-.08em\TeX}

\begin{document}

\pagestyle{fancy}
\fancyhead{}

\maketitle 


\section{Introduction}
\label{section:introduction}

A multi-agent system models the behaviour of a computing system by viewing it as the outcome of interactions between 
rational agents who can act independently or cooperate to achieve a goal \cite{Wooldridge2009}.
Studying a multi-agent system therefore means examining the strategic abilities of a group of agents \cite{MCMAC-LomuscioRaimondi2006,Jamroga15specificationMAS}.
In order to formally verify whether the executions produced by their interactions behave as expected,
strategic logics were introduced, augmenting temporal logics with strategy operators \cite{ATL02,Jamroga15specificationMAS,Jamroga18,PenczekAAMAS2023,BerthonMMRV17}.
They can express that a group of agents has a strategy to enforce a given temporal property,
regardless of the actions executed by the agents outside the group.
Formal verification using strategic logics addresses two main problems  \cite{Bulling10verification} stated as follows. 
Given a model of the system and a strategic formula:
\begin{enumerate}
    \item \textbf{Model checking:} determine whether the formula holds in the initial state of the model.
    \item \textbf{Strategy synthesis:} explicitly construct a strategy that ensures the temporal property is satisfied.
\end{enumerate}
A solution to the latter naturally yields a solution to the former.

Strategic Timed Computation Tree Logic (\STCTL)~\cite{PenczekAAMAS2023} is a powerful logic designed to describe the behaviour 
of systems whose executions are subject to explicit time constraints. %
It offers a natural framework for capturing scenarios in which the timing of actions is not merely 
incidental but plays a decisive role in determining whether a system meets its intended specification.
In this paper, we focus on the synthesis of memoryless strategies for agents having imperfect information about the system. This assumption is essential: in many real-world scenarios, agents do not have full visibility of the global state, and their decisions must rely solely on their observation of the system. Working under imperfect information therefore captures a more realistic view of strategic reasoning \cite{BerthonMMRV17}. 

\paragraph{Application Domain.}
Strategy synthesis is particularly valuable in safety-critical and time-sensitive multi-agent applications, such as industrial automation, avionics, and distributed control systems, where correctness and timely execution are paramount \cite{Rochange16,Singh24,WolfChapter9}.
Attack-defence trees \cite{KordyMRS10} representing relevant security scenarios can be easily translated to the multi-agent setting of our tool \cite{ADT2AMAS},
and studied in a new, strategic context of opposing coalitions \cite{ADTrees-ICFEM,ADTrees-ICECCS}.
Socio-technical interactions, \eg e-voting \cite{AIML2022} emerge as a use case.

\section{Comparison of Tools}
Reasoning about strategies in multi-agent systems has led to several verification tools, 
mostly focused on model checking. 
Notable examples include MCMAS \cite{Lomuscio15mcmas} and MCMAS-SLK \cite{CermakLMM14,CermakLMM18},
which support untimed strategic logics. 
Similarly, MOCHA \cite{MOCHA,TGC2} and STV \cite{STVtool-KurpiewskiJamrogaKnapik2019,AAMAS-KurpiewskiPJK21,STV+FLY} 
target strategic verification, including imperfect information, yet lack timing constraits.
Recent approaches leverage SMT for strategic logics. MsATL \cite{MSATL-AAMAS2020} checks ATL satisfiablity via SMT; 
SGSAT supports \SLSG \cite{Kacprzak-KR2021-SLSG}; and 
SMT4SMTL \cite{ECAI2023} enables model checking for SMTL \cite{ECAI2023}, 
a strategic timed logic based on LTL \cite{Pnueli77}.
The initial approach to \STCTL model checking encodes strategies using \emph{parameters}  \cite{PenczekAAMAS2023} in \imitator \cite{Andre2021imitator}, 
which is far from optimal.
Maude \cite{PPDP2024}, based on rewriting logic, can support model checking 
and strategy synthesis for full \STCTL.
It is not a dedicated model checker, requiring expert knowledge 
from the user to write the specification and optimally guide DFS or BFS exploration
(\imitator\ only uses the latter).
\tool\ natively supports strategic abilities for both model checking and 
synthesis of all strategies.
It advances these efforts by fully supporting \STCTL, 
and exhibits a dramatic speedup over \imitator\ and Maude.

\section{Theoretical Background}
\label{section:theoretical-background-problem-scenario}
In this section, we introduce the relevant theoretical background.

\paragraph*{Multi-agent systems.}
A continuous-time asynchronous multi-agent system (\TAMAS) \cite{PenczekAAMAS2023}
models agents as components of a timed automata network.
Clock constraints are enforced in transition guards and state invariants \cite{ALUR1992-TimedAutomata}.
Agents interleave private actions independently, but must synchronise on shared ones.
Their product (or \TAMAS \emph{model}) captures the system's global behaviour.

\paragraph*{Strategic ability.}
Conditional plans defining choices to be made by (groups of) agents are called (joint) \emph{strategies},
and classified by state information (perfect I vs. imperfect i)
and memory (perfect recall R vs. memoryless r) available to agents \cite{Schobbens04}.
We focus on \ir-strategies, \ie decisions based solely on the current local state.

\paragraph*{Logic.}
Strategic Timed \CTL\ (\STCTL)~\cite{PenczekAAMAS2023} extends \CTL\
by adding time constraints and strategic operators.
It is defined by the syntax:
\begin{center}
$\varphi ::= \prop{p} \mid \neg \varphi \mid \varphi\wedge\varphi \mid \coop{A}\gamma$,\\
$\gamma ::= \varphi \mid \neg\gamma \mid \gamma \wedge\gamma 
          \mid \Apath\gamma\Until_I\gamma \mid \Epath\gamma\Until_I\gamma \mid \Apath\gamma\Release_I\gamma \mid \Epath\gamma\Release_I\gamma$,
\end{center}
\noindent
where $\prop{p}$ is an atomic proposition, $\A$ is a subset of agents,
$\coop{\A}\gamma$ expresses that coalition $\A$ has a strategy to enforce the property $\gamma$,
and intervals $I\subseteq\mathbb{R}_{0+}$ denote time constraints on the evaluation of temporal operators.
The remaining \CTL\ path quantifiers ($\Apath,\Epath$), 
temporal operators ($\Until_I,\Release_I$, with $\Sometm_I,\Always_I$ derived), and 
Boolean connectives ($\neg,\wedge)$ are standard 
(see \cite{PenczekAAMAS2023} for their formal semantics).

\section{Architecture, Technology, and usage for strategy synthesis}
\label{section:architecture-and-technology}
The implementation of \tool\ extends the existing model checker \imitator \cite{Andre2021imitator}, 
written in OCaml.
\imitator is a state-of-the-art tool, supporting \TCTL\ properties over real-time systems 
represented as networks of Timed Automata (TA) with asynchronous execution.
Notably, it also enables \emph{parametric} verification, \ie Parametric TA input models,
where parameters representing constants with unknown values, may appear in the transition 
guards and local state invariants.
The nodes of the global state space, constructed via breadth-first search (BFS),
are therefore tuples of automata locations and constraints on parameters and clocks.

In the strategic context, the automata in the input model are 
thought of as autonomous \emph{agents}, which have their own behaviour and synchronise 
on shared actions. 
Their executed transitions are labelled by \emph{actions} selected as part of a \emph{strategy}.
To adapt \imitator to \STCTL\ model checking and strategy synthesis, global states are 
augmented with a representation of the current strategy being constructed.
During the BFS exploration, a successor can only be added if its strategy is compatible with 
the one already chosen for the source state, or if there is none yet 
(in which case, the strategy is updated for the target state).
Thus, the constructed state space
only contains the executions consistent with the strategy (called its \emph{outcome paths}).
The states that satisfy the \TCTL\ property inside the strategic operator are checked on-the-fly, 
and the tool displays the valid strategies and associated parameters\footnote{
Timing parameters synthesis is obtained here for free by extending \imitator.
}.
Note that this might not terminate but the partial results obtained are sound.

The \tool\ repository and companion artifact can be accessed at \url{https://hub.docker.com/r/imitator/aamas2026}, and the tool's demonstration video at \url{https://youtu.be/E1VpFxotNG0}.

\section{Experimental Evaluation}
\label{section:experimental-evaluation}
We compare the efficiency of \tool\ with the previous approach 
involving \imitator\ \cite{PenczekAAMAS2023}, on a benchmark of the voting example \cite{PenczekAAMAS2023}, though we additionally scale the size of the coalition.
The formula $\coop{V_i}\Epath\Sometm_{[0;8]}\prop{voted_{i,1}}$ specifies that voter $V_i$ has a strategy to vote for the first candidate within 8 time units.
In the experiments, we synthesise \emph{all} strategies
(by inputting the property with the \lstinline[basicstyle=\ttfamily]{#synth} keyword).
The results, summarised in \Cref{tab:results}, show huge 
gains (up to 35x) compared to the initial approach of \cite{PenczekAAMAS2023}.
All times are given in seconds, with the timeout set at 120s.

\begin{table}[h]
\centering
\begin{tabular}{ |c|c|c|c|c|c|c|c|c|c| }
 \hline
 \multicolumn{2}{|c|}{} & \multicolumn{4}{c|}{\imitator} & \multicolumn{4}{c|}{\tool} \\
 \hline
 |A| & v & c=1 & c=2 & c=3 & c=4 & c=1 & c=2 & c=3 & c=4 \\
 \hline
	1 & 1 & 0.01 & 0.01 & 0.02 & 0.03 & 0.01 & 0.01 & 0.01 & 0.01 \\
 \hline
	1 & 2 & 0.06 & 0.14 & 0.38 & 0.70 & 0.02 & 0.03 & 0.04 & 0.05 \\
 \hline
	1 & 3 & 0.67 & 2.15 & 6.49 & 16.5 & 0.10 & 0.21 & 0.44 & 0.81 \\
 \hline
	1 & 4 & 6.03 & 26.4 & 78.0 & timeout & 0.76 & 2.74 & 7.95 & 21.3 \\
 \hline
	1 & 5 & 45.4 & \multicolumn{3}{c|}{timeout} & 9.51 & 90.8 & \multicolumn{2}{c|}{timeout} \\
 \hline
	2 & 2 & 0.09 & 0.25 & 0.62 & 1.38 & 0.02 & 0.03 & 0.04 & 0.05 \\
 \hline
	2 & 3 & 0.92 & 3.55 & 10.0 & 27.5 & 0.12 & 0.27 & 0.53 & 1.02 \\
 \hline
	2 & 4 & 10.4 & 53.9 & \multicolumn{2}{c|}{timeout} & 1.01 & 3.75 & 12.0 & 34.0 \\
 \hline
	2 & 5 & 98.1 & \multicolumn{3}{c|}{timeout} & 15.5 & \multicolumn{3}{c|}{timeout} \\
 \hline
	3 & 3 & 1.21 & 5.16 & 17.2 & 47.2 & 0.12 & 0.29 & 0.65 & 1.33 \\
 \hline
	3 & 4 & 14.4 & 67.3 & \multicolumn{2}{c|}{timeout} & 1.18 & 4.72 & 16.4 & 53.7 \\
 \hline
	3 & 5 & \multicolumn{4}{c|}{timeout} & 20.3 & \multicolumn{3}{c|}{timeout} \\
 \hline
\end{tabular}
\caption{Synthesising all strategies in the voting benchmark with $v$ voters, $c$ candidates, and agent coalition of size $|\A|$.}
\label{tab:results}
\end{table}
\vspace{-0.6cm}

Moreover, the demonstration video features additional examples:
treasure hunters \cite{ADTrees-ICECCS} with added clocks,
specified as a Parametric TA to showcase also timing constraints synthesis,
and a variant of the conference scenario of \cite{Paradoxes-KR}.
We refer the reader to \cite[Table 1]{PPDP2024} for a suite 
of Parametric TA benchmarks from the library \cite{AndreMP21}, 
where \imitator\ (so \tool\ as well) outperforms Maude for many instances.
While these involve non-strategic properties, we note that \tool\ enables verification of the same models with more expressive specification involving strategic abilities.

\section{Conclusions and Future Work}
\label{section:conclusions}
\tool{} provides an integrated environment for synthesis of memoryless strategies with imperfect information, as a push-button procedure. 
The only input needed from the user are the model and the property. 
Compared to other approaches, it does not require additional expertise.
This tool constitutes a major step forward in strategy synthesis for real-time multi-agent systems.

An important direction for further extending \tool\ is adding support for other 
strategy semantics, \eg counting \cite{KnapikAPJP19}, timed,
or an intermediate ``partial view'' between imperfect and perfect information,
with strategic choices based on the local states of a \emph{subset} of agents
(not necessarily from the coalition).


\begin{acks}
This work was supported by:
CNRS IRP ``Le Tr\'{o}jk{\k a}t'',
the ANR-22-CE48-0012 project BISOUS,
the PHC Polonium project MoCcA (BPN/BFR/2023/1/00045).
We thank Jaime Arias for his help with the demonstration video and companion artifact of this paper.
\end{acks}


\balance
\bibliographystyle{ACM-Reference-Format} 
\bibliography{bibfile-IMITATOR4AMAS2025,bibfile-KR2025-onlycited}

@article{ATL02,
   author    = "Rajeev Alur and Thomas A. Henzinger and Orna Kupferman",
   title     = "Alternating-time Temporal Logic",
   journal   = "Journal of the ACM",
   volume    = "49(5)",
   pages     = "672--713",
   year      = "2002"
}

@InProceedings{ALUR1992-TimedAutomata,
	author="Alur, Rajeev
	and Dill, David",
	title="The Theory of Timed Automata",
	booktitle="Real-Time: Theory in Practice",
	year="1992",
	publisher="Springer",
	pages="45--73"
}

@incollection{Bulling10verification,
    author =      {N. Bulling and J. Dix and W. Jamroga},
    title     =   {Model Checking Logics of Strategic Ability: Complexity},
    booktitle =   {Specification and Verification of Multi-Agent Systems},
    publisher =   {Springer},
    pages =       {125--159},
    year =        {2010},
}

@book{Jamroga15specificationMAS,
    author    = "Wojciech Jamroga",
    title     = "Logical Methods for Specification and Verification of Multi-Agent Systems",
    publisher = "ICS PAS Publishing House",
    year      = "2015",
    isbn      = {978-83-63159-25-2},
}

@inproceedings{MCMAC-LomuscioRaimondi2006,
  author    = {Alessio Lomuscio and
               Franco Raimondi},
  title     = {Model Checking Knowledge, Strategies, and Games in Multi-Agent Systems},
  booktitle = {Proc. of the Int. Conf. on Autonomous Agents and Multi-Agent Systems (AAMAS'06)},
  pages     = {161--168},
  year      = {2006}
}

@inproceedings{MsATL-AAMAS2020,
    author    = {Artur Niewiadomski and
               Magdalena Kacprzak and
               Damian Kurpiewski and
               Michal Knapik and
               Wojciech Penczek and
               Wojciech Jamroga},
    title     = {{M}s{ATL}: {A} Tool for {SAT}- Based {ATL} Satisfiability Checking},
    booktitle = {Proceedings of {AAMAS'20}},
    pages     = {2111--2113},
    publisher = {IFAAMAS},
    year      = {2020}
}

@inproceedings{Pnueli77, 
	author = {Pnueli, Amir}, 
	title = {The Temporal Logic of Programs}, 
	publisher = {IEEE Computer Society}, 
	booktitle = {Proceedings of {SFCS '77}}, 
	pages = {46-57}, 
	numpages = {12},
	year = {1977}
}

@inproceedings{STVtool-KurpiewskiJamrogaKnapik2019,
  author    = {Damian Kurpiewski and
               Wojciech Jamroga and
               Michal Knapik},
  title     = {{STV:} Model Checking for Strategies under Imperfect Information},
  booktitle = {Proceedings of {AAMAS'19}},
  pages     = {2372--2374},
  year      = {2019}
}

@inproceedings{AAMAS-KurpiewskiPJK21,
author={Kurpiewski, Damian and Pazderski, Witold and Jamroga, Wojciech and Kim, Yan},
  title        = {STV+Reductions: Towards Practical Verification of Strategic Ability
                  Using Model Reductions},
  booktitle = {Proceedings of {AAMAS'21}},
  pages        = {1770--1772},
  publisher    = {{ACM}},
  year         = {2021}
}

@InProceedings{Schobbens04,
	author    = "Pierre-Yves Schobbens",
	title     = "{A}lternating-Time {L}ogic with {I}mperfect {R}ecall",
	booktitle = "Proceedings of {LCMAS} 2003",
	publisher = {Elsevier},
	year      = {2004},
	pages     = "1--12"
}

@article{Lomuscio15mcmas,
  author = {Alessio Lomuscio and Hongyang Qu and Franco Raimondi},
  title = {{MCMAS: An Open-Source Model Checker for the Verification of Multi-Agent Systems}},
  journal = {Int. J. Softw. Tools Technol. Transfer},
  volume = {24},
  pages = {84-90},
  publisher = {Springer},
  year = {2015}
}

@inproceedings{PPDP2024,
  author       = {Jaime Arias and
                  Carlos Olarte and
                  Wojciech Penczek and
                  Laure Petrucci and
                  Teofil Sidoruk},
  title        = {Model Checking and Synthesis for Strategic Timed {CTL} using Strategies
                  in Rewriting Logic},
  booktitle    = {Proceedings of {PPDP} 2024},
  pages        = {10:1--10:14},
  publisher    = {{ACM}},
  year         = {2024}
}

@inproceedings{KordyMRS10,
  author    = {Barbara Kordy
	and Sjouke Mauw
	and Sa$\check{\mathrm{s}}$a Radomirovi{\'{c}}
	and Patric Schweitzer},
  title     = {{F}oundations of {A}ttack-{D}efense {T}rees},
  booktitle = {Proceedings of {FAST} 2010},
  pages     = {80--95},
  year      = {2011},
  publisher = {Springer}
}

@inproceedings{ADTrees-ICFEM,
  author    = {Jaime Arias and Carlos Budde and Wojciech Penczek and Laure Petrucci and Teofil Sidoruk and Mari{\"{e}}lle Stoelinga},
  title     = {{H}ackers vs. {S}ecurity: {A}ttack-{D}efence {T}rees as {A}synchronous {M}ulti-agent {S}ystems},
  booktitle = {Proceedings of {ICFEM} 2020},
  pages     = {3--19},
  publisher = {Springer},
  year      = {2020}
}

@inproceedings{ADTrees-ICECCS,
  author    = {Laure Petrucci and Michal Knapik and Wojciech Penczek and Teofil Sidoruk},
  title     = {{S}queezing {S}tate {S}paces of ({A}ttack-{D}efence) {T}rees},
  booktitle = {Proceedings of {ICECCS} 2019},
  pages     = {71--80},
  publisher = {{IEEE}},
  year      = {2019}
}

@inproceedings{ADT2AMAS,
  author    = {Jaime Arias and Wojciech Penczek and Laure Petrucci and Teofil Sidoruk},
  title     = {{ADT2AMAS:} {M}anaging {A}gents in {A}ttack-{D}efence {S}cenarios},
  booktitle = {Proceedings of {AAMAS} '21},
  pages     = {1749--1751},
  publisher = {{ACM}},
  year      = {2021}
}

@inproceedings{Rochange16,
  author       = {Christine Rochange},
  title        = {Parallel Real-Time Tasks, as Viewed by {WCET} Analysis and Task Scheduling
                  Approaches},
  booktitle    = {Proceedings of {WCET} 2016},
  pages        = {1--11},
  publisher    = {Schloss Dagstuhl - Leibniz-Zentrum f{\"{u}}r Informatik},
  year         = {2016}
}

@article{Singh24,
  author       = {Abhishek Singh},
  title        = {Cutting-Plane Algorithms for Preemptive Uniprocessor Scheduling Problems},
  journal      = {Real Time Syst.},
  volume       = {60},
  number       = {1},
  pages        = {24--73},
  year         = {2024}
}

@incollection{WolfChapter9,
title = {Chapter 9 - Automotive and Aerospace Systems},
booktitle = {Computers as Components (Fifth Edition)},
publisher = {Morgan Kaufmann},
pages = {437-452},
year = {2023},
author = {Marilyn Wolf}
}

@inproceedings{AIML2022,
  author    = {Damian Kurpiewski and Wojciech Jamroga and {\L}ukasz Maśko and {\L}ukasz Mikulski and Witold Pazderski and Wojciech Penczek and Teofil Sidoruk},
  title     = {{V}erification of {M}ulti-{A}gent {P}roperties in {E}lectronic {V}oting: {A} {C}ase {S}tudy},
  booktitle = {Proceedings of AiML 2022},
  pages     = {531--556},
  publisher = {College Publications},
  year      = {2022}
}

@inproceedings{AndreMP21,
  author       = {{\'{E}}tienne Andr{\'{e}} and
                  Dylan Marinho and
                  Jaco van de Pol},
  title        = {A Benchmarks Library for Extended Parametric Timed Automata},
  booktitle    = {Proceedings of {TAP} 2021},
  pages        = {39--50},
  publisher    = {Springer},
  year         = {2021}
}

@inproceedings{Paradoxes-KR,
  author    = {Wojciech Jamroga and
               Wojciech Penczek and
               Teofil Sidoruk},
  title     = {Strategic Abilities of Asynchronous Agents: Semantic Side Effects
               and How to Tame Them},
  booktitle = {Proceedings of {KR} 2021},
  pages     = {368--378},
  year      = {2021}
}

@inproceedings{BerthonMMRV17,
  author       = {Rapha{\"{e}}l Berthon and
                  Bastien Maubert and
                  Aniello Murano and
                  Sasha Rubin and
                  Moshe Y. Vardi},
  title        = {Strategy logic with imperfect information},
  booktitle    = {Proceedings of {LICS} 2017},
  pages        = {1--12},
  publisher    = {{IEEE} Computer Society},
  year         = {2017}
}

@book{Wooldridge2009,
  author       = {Michael J. Wooldridge},
  title        = {An Introduction to MultiAgent Systems, Second Edition},
  publisher    = {Wiley},
  year         = {2009},
  isbn         = {978-0-470-51946-2}
}

@inproceedings{Jamroga18,
  author       = {Wojciech Jamroga},
  title        = {Model Checking Strategic Ability - Why, What, and Especially: How?},
  booktitle    = {Proceedings of {TIME} 2018},
  pages        = {1--10},
  publisher    = {Schloss Dagstuhl - Leibniz-Zentrum f{\"{u}}r Informatik},
  year         = {2018}
}

@inproceedings{STV+FLY,
  author       = {Damian Kurpiewski and
                  Mateusz Kaminski and
                  Wojciech Jamroga},
  title        = {{STV+FLY:} On-the-Fly Model Checking of Strategic Ability in Multi-Agent
                  Systems},
  booktitle    = {Proceedings of {ECAI} 2024},
  pages        = {4483--4486},
  publisher    = {{IOS} Press},
  year         = {2024}
}

@inproceedings{CermakLMM14,
  author       = {Petr Cerm{\'{a}}k and
                  Alessio Lomuscio and
                  Fabio Mogavero and
                  Aniello Murano},
  title        = {{MCMAS-SLK:} {A} Model Checker for the Verification of Strategy Logic
                  Specifications},
  booktitle    = {Proceedings of {CAV} 2014},
  pages        = {525--532},
  publisher    = {Springer},
  year         = {2014}
}

@InProceedings{MOCHA,
author="Alur, Rajeev
and Henzinger, Thomas. A.
and Mang, Freddy Y. C.
and Qadeer, Shaz
and Rajamani, Sriram K.
and Tasiran, Serdar",
title="MOCHA: Modularity in Model Checking",
year="1998",
booktitle="Proceedings of CAV 1998",
publisher="Springer",
pages="521--525",
isbn="978-3-540-69339-0"
}

@article{CermakLMM18,
  author       = {Petr Cerm{\'{a}}k and
                  Alessio Lomuscio and
                  Fabio Mogavero and
                  Aniello Murano},
  title        = {Practical Verification of Multi-agent Systems Against {SLK} Specifications},
  journal      = {Inf. Comput.},
  volume       = {261},
  pages        = {588--614},
  year         = {2018}
}

@inproceedings{PenczekAAMAS2023,
	author =        {Jaime Arias and Wojciech Jamroga and Wojciech Penczek and
	Laure Petrucci and Teofil Sidoruk},
	booktitle =     {Proceedings of {AAMAS}'23},
	pages =         {382--390},
	publisher =     {{ACM}},
	title =         {{S}trategic ({T}imed) {C}omputation {T}ree {L}ogic},
	year =          {2023}
}

@inproceedings{Andre2021imitator,
	title={IMITATOR 3: Synthesis of Timing Parameters Beyond Decidability},
	author={Andr{\'e}, {\'E}tienne},
	booktitle={Proceedings of CAV 2021},
	pages={552--565},
	year={2021},
	organization={Springer}
}

@inproceedings{ECAI2023,
  author =        {Magdalena Kacprzak and Artur Niewiadomski and
                   Wojciech Penczek and Andrzej Zbrzezny},
  booktitle =     {Proceedings of {ECAI} 2023},
  pages =         {1180--1189},
  publisher =     {{IOS} Press},
  title =         {{SMT}-Based Satisfiability Checking of Strategic Metric
                   Temporal Logic},
  year =          {2023},
}

@inproceedings{Kacprzak-KR2021-SLSG,
  author =        {Magdalena Kacprzak and Artur Niewiadomski and Wojciech Penczek},
  booktitle =     {Proceedings of {KR} 2021},
  pages =         {400--410},
  title =         {Satisfiability Checking of Strategy Logic with Simple
                   Goals},
  year =          {2021},
}

@article{KnapikAPJP19,
  author =        {Michal Knapik and {\'{E}}tienne Andr{\'{e}} and
                   Laure Petrucci and Wojciech Jamroga and
                   Wojciech Penczek},
  journal =       {J. Artif. Intell. Res.},
  pages =         {197--223},
  title =         {Timed {ATL:} Forget Memory, Just Count},
  volume =        {66},
  year =          {2019},
}

@inproceedings{TGC2,
author = {van der Hoek, Wiebe and Wooldridge, Michael},
title = {Tractable Multiagent Planning for Epistemic Goals},
year = {2002},
isbn = {1581134800},
publisher = {ACM},
booktitle = {Proceedings of {AAMAS'02}},
pages = {1167–1174}
}


\newpage
\nobalance
\onecolumn
\section*{Technical Requirements}
\label{section:requirements}
To replicate the experiments, install the latest Docker package and run the following commands (tested under Ubuntu 22.04 running via WSL, and on MacOS). The \lstinline[basicstyle=\ttfamily\small]{imitator/aamas2026} container should be automatically downloaded the first time \lstinline[basicstyle=\ttfamily\small]{docker run} is called.\\

\noindent
Voters example:

\lstinline[basicstyle=\ttfamily\small]{docker run imitator/aamas2026 /imitator/examples/voters.imi /imitator/examples/voters.imiprop -no-var-autoremove}\\

\noindent
Treasure hunters example:

\lstinline[basicstyle=\ttfamily\small]{docker run imitator/aamas2026 /imitator/examples/treasure_clocks.imi /imitator/examples/treasure.imiprop -no-var-autoremove}\\

\noindent
Conference example:

\lstinline[basicstyle=\ttfamily\small]{docker run imitator/aamas2026 /imitator/examples/conf2.imi /imitator/examples/conf2.imiprop -no-var-autoremove}\\

\noindent
Linux environment:

To access the source files of the tool and of the examples, you can run docker so as to work in a linux environment:

\lstinline[basicstyle=\ttfamily\small]{docker run --rm -it --entrypoint "/bin/bash" -v YOUR_LOCAL_EXAMPLES_DIRECTORY:/COPY_IN_THE_IMAGE -w /imitator imitator/aamas2026}\\

Where the following options are used:

\begin{description}
\item {\ttfamily\small --rm}: clean the container files when exiting
\item {\ttfamily\small -it}: interactive usage with a terminal window
\item {\ttfamily\small --entrypoint "/bin/bash"}: run the shell when starting the container instead of the imitator tool
\item{\ttfamily\small -v YOUR\_LOCAL\_EXAMPLES\_DIRECTORY:/COPY\_IN\_THE\_IMAGE}: if you want to make your own examples, you can link their folder on your own machine with a folder in the docker image that is run. You can then easily edit them as usual on your machine and test them in the docker image.
Then replace {\ttfamily\small YOUR\_LOCAL\_EXAMPLES\_DIRECTORY} with the path to your folder, and {\ttfamily\small COPY\_IN\_THE\_IMAGE} with the name in the docker. It will be found (because of the /) at the root of the directories tree.

If you don't plan to use this possibility, just remove this option from the command line.
\item {\ttfamily\small -w /imitator}: the working directory where you start. If you are using the previous option, you could write {\ttfamily\small -w /COPY\_IN\_THE\_IMAGE} to work directly in the folder of your own files.
\end{description}

The tool is then run, as in the video, using e.g.:

\lstinline[basicstyle=\ttfamily\small]{/imitator/bin/imitator /imitator/examples/treasure_clocks.imi /imitator/examples/treasure.imiprop -no-var-autoremove}\\

\end{document}